\documentstyle[aas2pp4]{article}
\def \cmsq           {\hbox{cm$^{-2}$}}
\def \deg          {\ifmmode ^{\circ}\else $^\circ$\fi}  
\def \etal         {{\it et~al.} }
\def \flam         {\hbox{ergs s$^{-1}$ cm$^{-2}$ \AA$^{-1}$}}    
\def \kms          {\rm{\hbox{km s$^{-1}$}}}
\def \lam          {$\lambda$}
\def \Lya          {\hbox{Ly$\alpha$}}

\def \Msun          {\hbox{M$_{\odot}$}}
\def \pcc           {\hbox{cm$^{-3}$}}
\def \zaz          {{$z_a\kern -1.5pt \approx\kern -1.5pt z_e$}}
\def \zllz         {{$z_a\kern -3pt \ll\kern -3pt z_e$}}
\def \Zsun          {\hbox{Z$_{\odot}$}}
\begin{document}

\renewcommand{\baselinestretch}{1.5}
\title
{\large\bf Broad P V Absorption in the BALQSO, PG~1254+047: \\
Column Densities, Ionizations and Metal Abundances in BAL Winds}
\renewcommand{\baselinestretch}{1}
\medskip

\author
{\large Fred Hamann}
\bigskip
\affil
{Center for Astrophysics and Space Sciences \\
University of California, San Diego; La Jolla, CA 92093-0424 \\
Internet: fhamann@ucsd.edu}
%
\renewcommand{\baselinestretch}{1}
\begin{abstract} 
\normalsize
This paper discusses the detection of 
\ion{P}{5}~\lam\lam 1118,1128 and other broad absorption lines 
(BALs) in archival {\it HST} spectra of the low-redshift BALQSO, 
PG~1254+047. The \ion{P}{5} identification 
is secured by excellent redshift and profile coincidences 
with the other BALs, such as \ion{C}{4}~\lam\lam 1548,1550 and 
\ion{Si}{4}~\lam\lam 1393,1403, and by photoionization calculations 
showing that other lines near this wavelength, 
e.g. \ion{Fe}{3}~\lam 1123, should be much weaker than \ion{P}{5}. 
The observed BAL strengths imply that 
either 1) there are extreme abundance ratios such as [C/H]~$\ga +1.0$, 
[Si/H]~$\ga +1.8$ and [P/C]~$\ga +2.2$, or 2) at least some of the lines 
are much more optically thick than they appear. 
\medskip

I argue that the significant presence of \ion{P}{5} absorption 
indicates severe line saturation, which is disguised in the observed 
(moderate-strength) BALs because the absorber does not fully cover the 
continuum source(s) along our line(s) of sight. The variety of observed BAL 
strengths and profiles results from a complex mixture of ionization, 
optical depth and coverage fraction effects, making useful determinations 
of the abundance ratios impossible without a specific physical model. 
Computed optical depths for all UV resonance lines show that the 
observed BALs are consistent with {\it solar} abundances 
if 1) the ionization parameter is at least moderately high, 
$\log U \ga -0.6$, 2) the total hydrogen 
column density is $\log N_{\rm H}$(\cmsq )~$\ga 22.0$, and 3) the optical 
depths in strong lines like \ion{C}{4} and \ion{O}{6}~\lam\lam 1032,1038 
are $\ga$25 and $\ga$80, respectively. These optical depths and column 
densities are at least an order of magnitude larger than expected from 
the residual intensities in the BAL troughs, but they are consistent 
with the large absorbing columns derived from X-ray observations of 
BALQSOs. In particular, a nominal X-ray column density of 
$\log N_{\rm H}$(\cmsq )~$\sim 23$ could produce the observed 
BAL spectrum if $+0.4\la \log U \la +0.7$ in a simple one-zone
medium. The outflowing BALR, at velocities from $-$15,000 to 
$-$27,000~\kms\ in PG~1254+047, is therefore a strong candidate 
for the X-ray absorber in BALQSOs. 

\end{abstract}

\keywords{Quasars: absorption lines; Quasars: general; 
Quasars: individual (PG~1254+047)}
\newpage

\section{Introduction}

Broad absorption lines (BALs) are prominent features in the UV spectra 
of $\sim$10 to 15\% of optically-selected (radio-quiet) QSOs 
(\cite{fol90,wey91}). The lines form in 
ionized winds where the speeds can range from 
near zero to more than 30,000~\kms\ (see \cite{wey95,tur95,tur88}, 
Weymann, Turnshek \& Christiansen 1985 for reviews). 
The most frequently measured lines are resonance transitions of 
\ion{H}{1} (\Lya~\lam1216) and moderate- to high-ionization metals  
such as \ion{C}{4}~\lam\lam 1548,1550, \ion{Si}{4}~\lam\lam 1393,1404, 
\\ \ion{N}{5}~\lam\lam 1239,1243 and 
\ion{O}{6}~\lam\lam 1032,1038. Lower ionization lines 
such as \ion{Mg}{2}~\lam\lam 2796,2804 and \ion{Al}{3}~\lam\lam 1855,1863 
are also detected in $\sim$15\% of BALQSOs in the optically-selected 
samples. 

One of the most surprising results from BAL studies is the 
extreme elemental abundances implied by the absorption-line column 
densities (\cite{wey85}, Junkkarinen, Burbidge \& Smith 1987, 
\cite{tur88}). Recent photoionization calculations 
that explore a wide range of ionization states and ionizing 
QSO spectral shapes (\cite{kor96,tur96,ham97d}) show, for 
example, that the measured ratios 
of \ion{Si}{4} to \ion{H}{1} column densities require typical {\it minimum}
Si/H abundances of $1.3\la {\rm [Si/H]}\la 2.0$, 
where [Si/H] is the logarithmic abundance ratio relative to 
solar, log~(Si/H)~$-$~log~(Si/H)$_{\odot}$. The surprising detections 
of broad  \ion{P}{5}~\lam\lam 1118,1128 absorption in 
PG~0946+301 (\cite{jun95,jun97}) and possibly Q~1413+113 
(\cite{tur88}) and Q~0226$-$104 (\cite{kor92}) 
indicate an even more extreme situation where phosphorus is highly 
overabundant, with [P/C]~$\ga 1.8$ and [P/H]~$\ga 3.0$ independent of 
the uncertain ionization state (see also \cite{ham97d} and \S6 below). 

Abundance determinations are important because they 
provide clues to the origin of the BAL gas, the physics of the acceleration 
mechanism and, perhaps, the nature of star formation and galactic nuclear  
evolution at high redshifts. 
Hamann \& Ferland (1993) and Hamann (1997) showed that gas-phase 
metallicities up to $\sim$10~\Zsun\ in QSOs 
are consistent with the vigorous star formation expected in the 
cores of young massive galaxies. However, the extreme abundances 
derived for the BAL regions (BALRs) require a more exotic enrichment 
scenario. Gas-phase metallicities above 10~\Zsun\ could, in principle, 
be reached by stellar populations with initial mass functions that are 
heavily weighted toward massive stars; but the overabundance of 
phosphorus is not compatible with any enrichment scheme dominated 
by Types I or II supernovae or by CNO-processed material from stellar 
envelopes. One possible solution, suggested 
by Shields (1996), is  that the BAL gas is enriched by dwarf 
novae. Shields (1996) proposed that the needed high nova rates 
(relative to supernova rates) might be achieved 
near QSOs by white dwarfs 
gaining mass as they plunge through the accretion disk that surrounds 
the black hole. The overall feasibility of the nova hypothesis has not been 
tested, but it is clear that the high metallicities and P/C ratios 
estimated for the BALR are at least consistent with nova 
abundances. 

Another possibility is that the BAL abundance results are simply 
incorrect. The abundance analysis hinges critically on the 
column densities derived from the residual intensities in 
BAL troughs. Typical estimates are 
$15.5 \la \log N$(\cmsq )~$\la 16.5$ for \ion{C}{4}, \ion{N}{5}, 
and \ion{O}{6}, with smaller values by roughly 0.5 dex for \ion{H}{1} 
and by $\sim$1.0 dex for \ion{Si}{4} 
(\cite{tur84,wey85,jun87,tur88,kor92,ham97d}). Estimates of the 
total hydrogen column densities (\ion{H}{1}+\ion{H}{2}) 
depend further on the uncertain ionization and/or metal abundances, 
but typical values are in the range 
$19\la \log N_{\rm H}$(\cmsq )~$\la 20$. 

It has long been recognized that these column densities are only lower 
limits (for example Junkkarinen \etal 1983 and 1987). 
High resolution spectroscopy has essentially ruled out 
the possibility that BAL troughs 
are composed of many unresolved components that ``hide'' 
large column densities (\cite{bar94}). However, resolved 
narrow and intermediate-width line components in a few BALQSOs 
(\cite{wam95,bar94}) reveal multiplet ratios that require large optical 
depths, even though the lines appear too shallow to be optically 
thick (also \cite{kor92}). 
Similar evidence has been noted in non-BAL (exclusively 
narrow) absorption-line systems that are also known or suspected of 
forming in QSO outflows 
(\cite{wam93,pet94,bar97b}, Barlow, Hamann \& Sargent 1997a, 
Hamann \etal 1997a, Hamann, Barlow \& Junkkarinen 1997b, and 
Hamann \etal 1997c). 
These results imply that the line-absorbing regions do not fully cover 
the background emission source(s); some of the flux emerges unabsorbed 
and fills in the bottoms of the absorption troughs. Large line 
optical depths and partial 
line-of-sight coverage of the emitting source(s) are also suggested 
by curiously similar line depths in different BAL troughs and  by 
BAL profiles that have nearly flat bottoms but 
fail to reach zero intensity (see \cite{ara97}). 

Similar conclusions have been drawn from spectropolarimetry, where 
higher percentage polarizations are often measured in 
BAL troughs compared to the adjacent continuum. Evidently, some 
of the continuum flux is not absorbed by the BAL 
gas -- perhaps because it is reflected into our line-of-sight by 
an extended scattering region (\cite{coh95,goo95,hin95}). 
The BALR might absorb the direct continuum radiation 
while the indirect (reflected) light is unabsorbed and thereby 
dilutes the observed BALs. 
The polarization and spectroscopic data therefore both indicate that 
the line optical depths and ionic column densities are larger 
than expected from the measured depths of the BALs. 

Consistent with these findings, 
theoretical analysis of low-ionization BAL systems (with \ion{Mg}{2} etc.) 
predicts that those BALRs should be optically thick at the 
Lyman edge (for roughly 
solar metallicities; \cite{voi93}). For a nominal BAL profile width 
of 10,000~\kms , \Lya\ will be $\sim$20 times more optically thick 
than the Lyman limit. Therefore, \Lya\ should be 
very saturated (in at least these low-ionization systems) even though it 
typically has a modest absorption depth in observed spectra (\cite{wey91}). 
This discrepancy between the observed and predicted \Lya\ troughs 
is probably due to partial line-of-sight coverage. 

Additional evidence for large absorbing columns 
has come from observations showing that BALQSOs 
are significantly weaker soft X-ray sources than non-BALQSOs 
with similar redshifts, luminosities and radio properties 
(\cite{kop94,gre96,gre97}). The 
spectral similarity of BAL and non-BAL QSOs in other bandpasses 
(e.g. \cite{wey91}) 
suggests that the difference in soft X-rays is caused by 
absorption rather than weaker emission in the BALQSOs. 
The implied X-ray absorbing columns are 
$\log N_{\rm H}$(\cmsq )~$\ga 22.3$ (for 
solar abundances; \cite{gre96}). In the two BALQSOs where soft X-rays 
are actually detected (\cite{sin87,mat95,gre96}), absorption is 
clearly indicated with a total column density of
$\log N_{\rm H}$(\cmsq )~$\sim 23.1$ in both cases. The ionization state 
of the X-ray absorber is unknown, but Mathur \etal (1995) and 
Green \& Mathur (1996) have suggested that it is 
similar to the highly-ionized (``warm'') absorbers measured in soft X-ray 
spectra of many Seyfert 1 galaxies and a few low-redshift non-BALQSOs. 
(see \cite{mat94a,mat94b,lao97,geo97,rey97} and references therein). 
If the high column density X-ray absorption in BALQSOs occurs within the 
BALR, the amount of high-velocity outflowing material would be
orders of magnitude larger than previous estimates based on the 
BALs alone.

This paper discusses the detection and implications of broad 
\ion{P}{5} absorption in a low-redshift
BALQSO, PG~1254+047 ($V=15.8$, $z_{em}=1.010$; \cite{hew93}). 
I argue that the detection of \ion{P}{5} absorption, and in 
particular its strength relative to \ion{Si}{4} and \ion{C}{4}, 
is not a signal of unusual abundances but rather of line optical 
depths and column densities that 
are at least ten times larger than the standard estimates. 

Sections 2-4 present the data and discuss the BAL profiles 
and identifications. Section 5 provides the standard analysis, 
with optical depths and column densities derived from the BAL troughs 
and photoionization calculations that constrain the ionization 
and metal abundances. This analysis leads to the relatively 
low column densities and extreme abundances described above. 
Section 6 reexamines the BAL data assuming the abundance ratios 
are roughly solar. That assumption leads to higher estimates of the 
ionization, column densities 
and line optical depths. Section 7 discusses some implications 
of these results and \S8 provides a summary. 

\section{The HST Archival Spectrum}

PG~1254+047 was observed with the {\it Hubble Space Telescope} 
({\it HST}) as part of the Absorption Line Key Project (\cite{bah93}) on 
17 February 1993. Spectra were obtained with the Faint Object 
Spectrograph (FOS) using the high-resolution gratings 
G190H (1590$-$2310~\AA ) and G270H (2220$-$3275~\AA ) and the 
$0.^{\prime\prime}25\times 2.^{\prime\prime}0$ entrance 
aperture. The spectral resolution with both gratings is roughly 
230~\kms\ across four pixels. (Note that, although these observations 
occurred before installation of the image corrector COSTAR, the small 
aperture prevents significant loss of spectral resolution.)
Four exposures with G190H and one with G270H provided total 
exposure times of 9160 and 1992 seconds, respectively. 
The individual spectra were calibrated and archived by the {\it Space 
Telecope Science Institute}. I shifted the wavelength scales slightly 
($<$2.0~\AA ) using the Galactic \ion{Mg}{2} absorption lines 
in the G270H spectra and sharp features common to both the 
G270H and G190H data. 

Figure 1 shows the combined {\it HST} spectrum derived by averaging 
the individual exposures\footnote{These manipulations 
and the fits to the data in \S3 were performed with the 
IRAF software, which is provided by the National 
Optical Astronomy Observatories under contract to the National Science 
Foundation.}. 

\section{BAL Profiles}

The dotted curve in Figure 1 is a low-order polynomial fit to the 
continuum, approximately a $\lambda^{-1.5}$ power law. The fit is 
constrained by the measured flux in 4 narrow 
wavelength bands, 1980-2030, 2145-2170, 2920-2970 and 
3190-3280~\AA . The fitted continuum is forced to lie slightly below 
the observed 3190-3280~\AA\ flux because weak broad emission lines 
(BELs) of \ion{He}{2}~\lam 1640 and \ion{O}{3}]~\lam 1664 probably 
contribute at those wavelengths. The fit ignores the numerous intervening 
(narrow) absorption lines in the 1980-2030~\AA\ band, and it is 
unconstrained and therefore highly 
uncertain at wavelengths below 1980~\AA\ ($\la$985~\AA\ rest). 
Spectra of non-BALQSOs appear to 
change slope between roughly 1000 and 1200~\AA\ and rise 
less steeply toward shorter wavelengths than the fit in Figure 1 
(\cite{zhe97}). Therefore, the extrapolation of this 
low-order polynomial probably overestimates the true continuum flux at 
wavelengths below the \ion{O}{6} BEL.  

The final fit in Figure 1 also includes an estimate of the 
\ion{O}{6} emission line, which appears to be partially absorbed by the 
\ion{P}{5} BAL. The synthesized \ion{O}{6} line has the same 
redshift and profile as the measured 
\ion{C}{4} BEL, while its strength is constrained 
loosely by the measured flux above the fitted continuum on the blue 
side of the \ion{O}{6} emission profile. The resulting ``fit'' to the 
\ion{O}{6} BEL is obviously uncertain, but its strength and profile relative 
to the other BELs are typical of low-redshift 
non-BALQSOs (\cite{lao94,lao95}). 

Figure 2 compares several of the line profiles to that of \ion{C}{4}, 
after normalization by the fit just described. 
The BALs are ``detached'' from the emission lines, 
appearing at velocities between roughly 
$-$27,000 and $-$15,000~\kms\ relative to the emission redshift. 
There is unrelated absorption at more negative velocities in the 
\ion{N}{5} and \ion{O}{6} profiles caused by blends with other lines 
(see \S4 below). The likely overestimate of the continuum flux across 
the \ion{O}{6} BAL probably also leads to an overestimate of the 
absorption strength in this profile. 

\section{BAL Indentifications}

Numerous candidate BALs are labeled in Figure 1 at redshifts 
corresponding to the 3 deepest minima in the 
\ion{C}{4}~\lam\lam 1548,1551 trough (see Fig. 2). 
Not all of the labeled transitions are 
detected. Broad absorption is clearly present in \ion{C}{4}, 
\ion{Si}{4}~\lam\lam 1394,1403, \ion{N}{5}~\lam\lam 1239,1243, 
\ion{P}{5}~\lam\lam 1118,1128, \ion{O}{6}~\lam\lam 1031,1038 
and probably in \Lya , 
\ion{C}{3}~\lam 977 and \ion{S}{6}~\lam\lam 933,945. Notably absent 
are significant BALs in the singly-ionized lines, \ion{Al}{2}~\lam 1671, 
\ion{C}{2}{~\lam 1335 and \ion{Si}{2}~\lam 1260. 
Ground-based spectra of PG~1254+047 
(\cite{ste91}) show that \ion{Mg}{2}~\lam\lam 2796,2804 and 
\ion{Al}{3}~\lam\lam 1855,1863 are also not significantly present. 
This source is therefore among the majority 
of so-called high-ionization (non-\ion{Mg}{2}) BALQSOs. 
The strengths, profiles and velocities of the BALs in 
PG~1254+047 are not in any way peculiar for this class of 
absorber (\cite{jun87,tur88,wey91,kor92}). 

The normality of this BAL system makes the \ion{P}{5} 
detection particularly interesting. 
The identification with \ion{P}{5} is supported by the
excellent redshift and profile coincidences with other BALs 
(Figs. 1 and 2). The only other transition that 
might compete with \ion{P}{5} near this wavelength is a resonance 
multiplet of \ion{Fe}{3} (UV1),
whose ground state transition at 1122.5~\AA\ lies between 
the \ion{P}{5} doublet members. In the only previous secure measurement 
of a \ion{P}{5} BAL (PG~0946+301; \cite{jun97}), 
the \ion{P}{5} identification is 
strongly favored over \ion{Fe}{3} because narrow line components provide 
tight constraints on the expected absorption profiles. In the only known 
non-BAL system with \ion{P}{5} absorption (an intrinsic \zaz\ system 
in Q~0449$-$135; Barlow \etal 1997a and 1997b), 
the much narrower lines confirm the \ion{P}{5} identification and rule 
out any contribution from \ion{Fe}{3}. The analogy with these sources 
suggests that \ion{P}{5} is also the correct identification for this 
feature in PG~1254+047. 

However, we can test the \ion{P}{5} identifications further by examining 
the theoretical BAL strengths discussed in \S5.2 and \S6 below. 
Figures 4 and 6 (see below) show predicted line optical 
depths for BALRs that have solar abundances and are in 
photoionization equilibrium with a standard QSO radiation field. 
The two plots differ in that Figure 4 assumes the absorber is optically 
thin in the continuum at all UV wavelengths, while Figure 6 includes 
significant column densities and therefore continuum opacities. 
The results are plotted for a range of ionization 
parameters, $U$ (defined as the dimensionless ratio of the density in 
hydrogen-ionizing photons to the total hydrogen particle density at 
the illuminated face of the clouds), where larger $U$ values produce 
higher ionization states (see \S5.2 and \S6 for details). 
The plots are designed to show which lines 
should appear with \ion{P}{5} for roughly solar abundance ratios. 
However, they also show that \ion{Fe}{3}~\lam 1123 
(in the middle panel of both figures) 
is much weaker than the numerous low-ionization lines 
that are not detected in PG~1254+047, such as \ion{Al}{2}~\lam 1671, 
\ion{Al}{3}~\lam\lam 1855,1863, \ion{Mg}{2}~\lam\lam 2796,2804 and 
\ion{C}{2}~\lam 1335. In fact, 
there is no range in $U$ where \ion{Fe}{3}~\lam 1123 
is stronger than the \ion{Al}{2}, \ion{Mg}{2} and \ion{C}{2} lines. 
Significant absorption in \ion{Fe}{3} is therefore effectively ruled out 
by the absence of the low-ionization lines. Above the 
ionization limits established below, 
$\log U \ga -1.8$ in Figure 4 and $\log U \ga -0.6$ in Figure 6, 
\ion{Fe}{3}~\lam 1123 
is at least $\sim$2.7 dex weaker than both \ion{P}{5}~\lam 1118,1128 
and the undetected low-ionization lines. 
Therefore, \ion{Fe}{3} cannot contribute to the 
absorption near 1120~\AA\ without extreme enhancements in the 
iron abundance. We conclude that \ion{P}{5} is the only viable 
identification for this BAL.

\section{Standard Analysis}

\subsection{Line Optical Depths and Ionic Column Densities}

Here I estimate the line optical depths, $\tau$, from the residual 
intensities, $I_r$, in the normalized BAL troughs (Fig. 2) using the relation 
$\tau = -\ln I_r$. Optical depths derived in this way for doublets such as 
\ion{C}{4} include contributions from both transitions at each velocity. 
The measured \ion{Si}{4} and possibly \ion{P}{5} BALs have 
absorption structure that corresponds roughly to the doublet 
separations in these lines (Fig. 2). I 
therefore use the procedure described by Junkkarinen \etal (1983) to 
remove the doublet structure and simulate the true run of optical 
depth versus velocity for single transitions. 
These corrected optical depth profiles are 
plotted in Figure 3 for \ion{C}{4}, \ion{Si}{4} and 
\ion{P}{5}. They are directly 
proportional to the amount of absorbing material (the column 
density) at each velocity. 

Table 1 lists the column densities in 
\ion{C}{4}, \ion{Si}{4} and \ion{P}{5} 
derived by integrating the optical depth profiles in Figure 3 
(with oscillator strengths from \cite{ver96}) . 
The photon counting statistics and uncertain
continuum placements lead to 1$\sigma$ uncertainties in 
these column densities of order 0.07 dex for \ion{C}{4} and \ion{Si}{4} 
and $\sim$0.17 dex for \ion{P}{5}.  
Table 1 also provides estimates of the column densities in \ion{H}{1} 
and \ion{N}{5} derived by scaling the optical depth profiles 
in \ion{Si}{4} and \ion{C}{4} to match the 
\ion{H}{1} and \ion{N}{5} BALs, respectively. The \ion{C}{4} 
optical depths are scaled to fit the observed \ion{N}{5} absorption at 
velocities between roughly $-$15,500 and $-$25,000~\kms , where the 
measured \ion{N}{5} profile should not be seriously 
contaminated by blends. The scaled \ion{C}{4} profile is then integrated 
to yield a column density for \ion{N}{5}. Similarly, the \ion{Si}{4} 
optical depth profile is scaled to fit the measured absorption in 
\Lya\ near $-$24,000~\kms . The resulting estimate of the 
\ion{H}{1} column density 
is only an upper limit because the scaling ignores possible 
contributions from overlapping lines such as \ion{Si}{3}. 
The decision to match the optical depths in \ion{N}{5} with 
\ion{C}{4} and in \Lya\ with \ion{Si}{4} comes from the assumption 
that BALs with more similar ionizations have more similar profiles. 
The 1$\sigma$ uncertainties in the \ion{H}{1} and 
\ion{N}{5} column densities should be $\la$0.2 dex. Note that 
the upper limit on the \ion{H}{1} column density is consistent with the 
absence of a strong edge at the Lyman limit (Fig. 1), where optical 
depth unity would require $\log N_{\rm HI}$(\cmsq )~$\approx 17.2$.

\subsection{Ionization and Total Column Density}

The column densities in Table 1 imply that the 
BALR is optically thin in the Lyman continuum out to at least 
the \ion{N}{5} ionization threshold of 98~eV (see \cite{ost89} 
for bound-free absorption cross-sections). 
It is therefore appropriate to compare these column densities to 
theoretical simulations of an optically thin absorbing medium. 
I use the numerical code CLOUDY (\cite{fer96}) to model absorbing 
regions that 1) are dust-free, 2) have a constant gas density of 
$n_{\rm H} = 10^8$~\pcc , 3) have solar element abundances, 
4) are optically thin throughout the Lyman continuum, and 5) are in 
photoionization equilibrium with a standard QSO spectrum. 
The ionizing spectrum is a piecewise power law 
($f_{\nu}\propto \nu^{\alpha}$) with $\alpha = -0.5$ for 
$0.125 \leq h\nu < 25$~eV, $\alpha = -2.1$ for 
$25 \leq h\nu < 700$~eV, and $\alpha = -1.0$ for $h\nu \geq 700$~eV.
The two-point spectral index between 2500~\AA\ and 2~keV is 
$\alpha_{ox} = -1.5$. This continuum shape is C(1.4,$-$1.5) in the 
notation of Hamann (1997); it is believed to be appropriate for 
bright, low-redshift QSOs like PG~1254+047\footnote{I do not 
adopt the measured limit of $\alpha_{ox} \leq -1.7$ for PG~1254+047 
itself (\cite{wil94}) because I will attribute the weak X-ray flux to 
absorption (\S7), consistent with other BALQSOs (\S1). The 
X-ray flux is of little consequence in these calculations anyway, because 
it has almost no effect on the relative abundances of low- to 
moderate-ionization species.} (but see \cite{zhe97} 
and Korista, Ferland \& Baldwin 1997 for discussion). None of the 
important conclusions would be altered by reasonable 
changes to the ionizing spectrum (see \cite{ham97d}). Also, 
the calculated results of interest here, 
namely, the ionization fractions at each $U$, 
are not sensitive to the density or metal abundances. Therefore, 
one can use these calculations to derive the abundances in 
non-solar abundance situations.

Figure 4 shows the calculated line optical 
depths relative to \ion{P}{5}~\lam 1118 for all important resonance lines 
between 912 and 3000~\AA\ (plus \ion{Ne}{8}~\lam\lam 770,780 but 
excepting lines in the higher \ion{H}{1} Lyman series). 
The oscillator strengths for these calculations are 
from Verner, Barthel \& Tytler (1994). 
The optical depths are derived assuming that 
the bound electrons are entirely in their 
ground states. This is a good approximation for most ions where 
the first excited levels are at least several eV above ground. 
However, a few ions have low-energy excited 
states that could be significantly populated. In those cases, 
for example \ion{C}{2}, \ion{Si}{2}, \ion{P}{3} and \ion{Fe}{3}, 
the ground and excited state transitions within a given multiplet 
have similar wavelengths and will be blended in BAL troughs.  
Therefore, the single-line results in Figure 4 remain good approximations  
to the total optical depths in the blended multiplets. (For example, 
the total optical depth in \ion{C}{2}~\lam 1335 + \ion{C}{2}$^*$~\lam 1336 
is well approximated by \ion{C}{2}~\lam 1335 in the figure.) 

The most robust, abundance-independent constraints on the ionization 
come from comparing the measured BALs in different ions of the same 
element. For PG~1254+047, we can compare \ion{C}{4} 
with \ion{C}{3} and \ion{N}{5} with \ion{N}{3}. 
These comparisons are hampered by the line blending and uncertain 
continuum location at short wavelengths (\S3). 
Nonetheless, it is clear from Figure 1 that the \ion{C}{4} absorption is 
stronger (has a larger optical depth) than \ion{C}{3}~\lam 977 and the 
\ion{N}{5} line is at least several times stronger 
than \ion{N}{3}~\lam 990 (which might be absent altogether). 
Figure 4 shows that this situation requires a minimum 
ionization parameter of $\log U\ga -1.8$. If the \ion{C}{3} BAL is 
correctly identified, its minimum strength relative to 
\ion{C}{4} places an upper limit on the ionization. I estimate 
that the optical depth in \ion{C}{3} is at least $\sim$0.2 
times that in \ion{C}{4}, which implies $\log U\la -1.2$.  

These ionization results are consistent with the relative strengths of 
the BALs of different elements. For example, inspection of Figures 1 and 
2 indicates that the optical depths in the \ion{N}{5} and \ion{O}{6} troughs  
are roughly 2 to 5 times larger than those in \ion{Si}{4}. 
According to Figure 4, this situation 
requires $-1.7\la\log U\la -1.5$ for solar abundance ratios among these 
elements. Similarly, I estimate that the missing low-ionization BALs,  
\ion{Mg}{2} and \ion{Al}{3} (\cite{ste91}), have maximum optical depths 
of $<$0.1 if their absorption profiles are similar to \ion{Si}{4} or 
\ion{C}{4}. The weakness of the low-ionization lines relative to 
\ion{N}{5} and \ion{O}{6} therefore implies $\log U \ga -2.0$. 

The simulations used to create Figure 4 also show that the 
ionization limits inferred from \ion{C}{3}/\ion{C}{4}, namely 
$-1.8\la \log U \la -1.2$, correspond to neutral hydrogen fractions, 
$f$(\ion{H}{1})~$\equiv N_{\rm HI}/N_{\rm H}$,  in the range 
$-4.0\la \log f({\rm HI})\la -3.4$ (see also Figure 2b in \cite{ham97d}). 
The upper limit on the \ion{H}{1} 
column density in Table 1 therefore implies a {\it maximum} 
column density in total hydrogen of 
$18.4\la \log N_{\rm H}$(\cmsq )~$\la 19.0$, independent of the 
metal abundances. 

\subsection{Heavy Element Abundances}

The relative abundance of any two elements $a$ and $b$ can be 
derived from the following equation, 
\begin{eqnarray}
\left[{a\over b}\right]  = \
\log\left({{N(a_i)}\over{N(b_j)}}\right)  +
\log\left({{f(b_j)}\over{f(a_i)}}\right)  +
\log\left({b\over a}\right)_{\odot} \nonumber
\end{eqnarray}
\begin{equation}
\equiv \ \log\left({{N(a_i)}\over{N(b_j)}}\right)  + IC
\end{equation}
where $(b/a)_{\odot}$ is the solar abundance ratio, and $N$ and $f$ 
are  respectively the column densities and ionization fractions 
of elements $a$ and $b$ in ion stages $i$ and $j$. $IC$ is the normalized 
ionization correction factor. Hamann (1997) presented values of $IC$ for 
a wide range of photoionization conditions in absorbing regions near QSOs. 
I showed that for optically thin absorbers there 
are minimum values of $IC$ that can be used to derive 
minimum metal-to-hydrogen abundance ratios even if there are 
no constraints on the ionization (see Figures 7-9 in 
that paper). I also showed that one can make conservatively low 
(but not quite minimum) estimates of the metal-to-hydrogen ratios 
by assuming that each metal line forms where that ion is most favored, 
ie. at $U$ values where the $f(a_i)$ curves are at their peaks (Figure 
2b in \cite{ham97d}). This assumption is 
most appropriate if there is a range of ionization states 
caused by a range of densities or distances from the continuum source. 

The last two columns in Table 1 give the logarithmic metal-to-hydrogen
abundance ratios derived from the measured column densities via 
Eqn. 1. All of these ratios are lower limits because they 
are based on an upper limit to the \ion{H}{1} column. 
The results listed for [M/H]$_{min}$ are the absolute minima  
for each metal M that follow from the minimum $IC$ values 
(from Figure 9 in \cite{ham97d}). 
The results labeled [M/H]$_p$ are the conservatively 
low ratios derived with the assumption that the $f(M_i)$ curves are 
at their peaks (with $IC$ from Figure 6 in \cite{ham97d}). 
The uncertain shape of the ionizing spectrum causes approximate 
1$\sigma$ uncertainties in the abundances of 
$\pm$0.15 dex (not including the uncertainties in the 
column densities themselves; see \cite{ham97d} for discussion). 

The same techniques can be used to derive conservatively low, 
and in some cases minimum, metal-to-metal abundance ratios. 
Figure 5 shows the values of $IC$ needed to derive 
[P/C], [P/Si] or [Si/C] abundances from Eqn. 1 given the measured 
ratios of \ion{P}{5}/\ion{C}{4}, \ion{P}{5}/\ion{Si}{4}, 
\ion{P}{4}/\ion{Si}{4} or \ion{Si}{4}/\ion{C}{4} 
column densities. The correction factors in this plot come from 
the same photoionization calculations used for Figure 4 (\S5.2). 
We can place a robust lower limit on [P/C] because 
the \ion{P}{5}/\ion{C}{4} correction factor has a firm minimum value 
of $\sim$3.1 (for any ionization state and any nominal continuum 
shape; see also Fig. 10 in Hamann 1997). 
This limit is [P/C]~$\ga 2.2$ for the column densities in Table 1. 
In contrast, the ionization corrections for \ion{P}{5}/\ion{Si}{4} and 
\ion{Si}{4}/\ion{C}{4} depend sharply on $U$. Above the 
ionization limit noted above, $\log U \ga -1.8$, I estimate 
[P/Si]~$\la 1.8$ and [Si/C]~$\ga 0.5$. 

\section{Line Saturation: Revised Estimates of the Physical Parameters}

The extreme abundances derived in the previous section might 
be a signal that something is wrong with that analysis. 
It is already evident from Figure 4 that the variety of BALs observed 
in PG~1254+047 would be consistent with {\it solar} abundances if at 
least some of the lines are more optically thick than they appear 
in the spectrum. The 
line strengths relative to \ion{P}{5} provide a test of the saturation 
hypothesis. In particular, for ionizations above $\log U \ga -1.6$ in 
Figure 4 there are no BALs clearly absent from PG~1254+047 
that have predicted optical depths larger than the \ion{P}{5} lines. 
Conversely, all of the lines with predicted optical depths larger 
than \ion{P}{5} are (or could be) significantly present in the 
PG~1254+047 spectrum. 

If the BALs are saturated, the optical depths 
and column densities derived in \S5.1 and \S5.2 are only lower limits 
and the abundance estimates in \S5.3 should be disregarded. 
The abundance results are corrupted not only 
by the uncertain column densities, but also by 
the use of photoionization calculations that assume 
the absorbing region is optically thin in the UV continuum. 
That assumption might be incorrect. The larger column densities 
implied by optically thick lines can produce significant 
continuous (bound-free) absorption in the BALR which, in turn, leads to 
a range of ionization states at each $U$ (recall that $U$ applies 
only to the illuminated face of the clouds). 
Experiments with CLOUDY show that the results in Figure 4 
are accurate for total column densities up to 
$\log N_{\rm H}$(\cmsq )~$\approx 20.5$; above that, one must examine 
simulations for particular choices of $N_{\rm H}$. 

Figure 6 shows the line optical depths calculated for constant density, 
photoionized clouds with $N_{\rm H}$ (labeled across the top of the 
figure) determined at each $U$ by the requirement that 
the \ion{P}{5} BAL (\lam 1118+\lam 1128) 
has optical depth 0.2. This \ion{P}{5} optical depth is an approximate 
average over the $\tau_{\rm PV}$ profile in Figure 3 and corresponds to 
the minimum column density of $\log N_{\rm PV} ({\rm\cmsq}) \sim 15.0$ 
in Table 1. The results plotted in Figure 6 are true optical depths 
(not ratios) for solar abundances and square absorption profiles with 
FWHM~=~10,000~\kms . Note that total column densities of at 
least $\log N_{\rm H}$(\cmsq )~$\ga 22.0 -$[P/H] 
are required to reach $\tau_{\rm PV} = 0.2$. 
Also, simulations with $\log U < -1.0$ are not plotted
because they have \ion{P}{5} optical depths well below 0.2 
for any $N_{\rm H}$. 

Figure 6 supports the claim that the observed 
assortment of BALs is consistent with solar abundances if at least 
some of the lines are optically thick. The observed 
\ion{C}{3}/\ion{C}{4} absorption strengths (\S5.2) now require 
a minimum ionization parameter of $\log U\ga -0.9$, which 
is also the minimum needed to produce $\tau_{\rm PV} \geq 0.2$. 
The upper limit on the \ion{Mg}{2} absorption, $\tau_{\rm MgII} < 0.1$ 
(\S5.2), relative to \ion{P}{5} 
implies $\log U\ga -0.6$ for roughly solar abundance ratios. 
If the ionization is high enough, the \ion{Si}{4} line might be 
optically thin with a column density close to the lower limit in Table 1. 
However, other minimum optical depths implied by Figure 6 are $\ga$6 
in \Lya , $\ga$25 in \ion{C}{4}, $\ga$12 in \ion{N}{5} and 
$\ga$80 in \ion{O}{6}, which correspond to minimum ionic column densities 
of $\log N_{\rm HI}$(\cmsq )~$\ga 16.7 -$[P/H],
$\log N_{\rm CIV}$(\cmsq )~$\ga 17.3 -$[P/C], 
$\log N_{\rm NV}$(\cmsq )~$\ga 17.2 -$[P/N], and 
$\log N_{\rm OVI}$(\cmsq )~$\ga 18.1 -$[P/O]. The scaling of 
these column densities with the abundance ratios is only approximate 
because each ion affects the radiative transfer and ionization balance 
in a different way. Nonetheless, if the abundance ratios are close to solar, 
the column densities in these ions must be at least an order of 
magnitude larger than the estimates in Table 1. 

If the \ion{P}{5} optical depth is larger than the minimum value of 
0.2, the $N_{\rm H}$ values in Figure 6 are only lower limits. Adding to 
the total column densities in those simulations would increase the optical 
depths in low-ionization lines, such as \ion{Mg}{2} and \ion{Al}{3}, 
but leave the high-ionization 
lines like \ion{N}{5} and \ion{O}{6} unchanged. Since \ion{Mg}{2} and 
\ion{Al}{3} are the strongest of the low-ionization lines, 
the upper limits on their optical depths in PG~1254+047 
define an upper limit on the total column 
density at each $U$. My estimate of $\tau < 0.1$ for both \ion{Mg}{2} and 
\ion{Al}{3} (\S5.2) corresponds to maximum column densities of 
$\log N_{\rm MgII}$(\cmsq )~$< 14.2$ and 
$\log N_{\rm AlIII}$(\cmsq )~$< 14.4$ for square absorption profiles 
with FWHM~=~10,000~\kms . Figure 7 compares the envelope of 
$N_{\rm H}$ upper limits defined by the maximum \ion{Mg}{2} and \ion{Al}{3} 
optical depths to the lower limits defined by $\tau_{\rm PV} \geq 0.2$ 
(from Fig. 6). The range of permitted $N_{\rm H}$ values lies between the 
two solid curves in Figure 7. The corresponding limits on $N_{\rm HI}$ 
are shown by the two dash-dot curves in Figure 7. Keeping in mind that 
$\log N_{\rm HI}$(\cmsq )~$\approx 17.2$ corresponds to optical depth 
unity at the Lyman limit (\cite{ost89}), the dash-dot curves indicate that 
this high-ionization absorber is in all cases optically thin at the Lyman 
edge (see also \cite{voi93}).

Obviously, these estimates of the hydrogen column densities depend on 
the assumption of solar metallicities and, in particular, on solar P/H, 
Mg/H and Al/H  ratios. Higher metal abundances would shift all of the 
curves in Figure 7 downward. Also, if the line optical depths {\it are} 
greater than expected from the depths of the BALs, then the upper 
limit on the \ion{Mg}{2} and \ion{Al}{3} optical depths could be larger 
than 0.1 and 
the upper limits on $N_{\rm H}$ and $N_{\rm HI}$ could be larger than 
they appear in Figure 7. The only firm upper limit on $N_{\rm H}$ comes 
from the requirement that the absorber is not too 
optically thick to electron scattering, 
$\log N_{\rm H}$(\cmsq )~$\la 24.2$ (but see \cite{goo97}). 

\section{Discussion}

The most surprising results from the standard analysis (\S5) 
are the extreme abundance ratios, 
[C/H]~$\ga +1.0$, [Si/H]~$\ga +1.8$ and [P/C]~$\ga +2.2$. 
However, I have shown (\S6) that the observed BALs are consistent 
with {\it solar} metal abundances if at least some of the lines 
are more optically thick than they appear. This interpretation is 
supported by the growing evidence for severe saturation in 
both the broad- and narrow-line intrinsic systems of other QSOs (\S1). 
The large line optical depths are masked in these systems by incomplete 
line-of-sight coverage of the background emission source(s). 
Resolved multiplets in narrow-line absorbers can provide 
both the coverage fractions and line optical depths 
(\cite{ham97a,bar97b}). It is therefore significant that 
the only known narrow-line system with \ion{P}{5} absorption also has  
(resolved) doublet ratios in \ion{C}{4}, \ion{Si}{4} and \ion{N}{5} that 
require line saturation and partial line-of-sight coverage (the \ion{P}{5} 
ratio could not be determined; 
Barlow \etal 1997a and 1997b). In BALs like PG~1254+047, 
multiplet ratios are not available and the optical depths and 
column densities inferred from the absorption troughs must be 
treated as lower limits. I propose that the significant presence of 
\ion{P}{5}~\lam\lam 1118,1128 absorption indicates that the optical 
depths and column densities are much larger than these limits. 

One result of the narrow-line studies is that different lines/ions 
can have different coverage fractions (\cite{ham97a,bar97a,bar97b}). 
No clear picture of the absorbing environments has yet emerged, but 
the coverage fractions at each profile velocity might depend not only 
on the geometry but also on the ionization and line optical depths. 
For example, low-ionization lines might generally have 
smaller coverage fractions because they 
form in relatively small and dense condensations within the BALR. 
Also, even if the BALR has just one ionization state (no dense 
condensations), different column densities along different 
sightlines could lead to larger {\it effective} coverage fractions 
in lines that are more optically thick. (Lines of sight through 
high-column density regions could be optically thick in many 
transitions -- giving rise to the \ion{P}{5} absorption and the overall 
saturated appearance of the spectrum, while sightlines through 
low-column density gas could be optically thick in just the strongest 
transitions -- giving them larger effective covering.) 
Some BALs might be weak simply because they are optically thin. 
However, coverage fractions effects can also 
produce differences in the BAL strengths and profiles,  
even among lines that are optically thick. These 
effects might naturally explain why 
low ionization BALs like \Lya\ and \ion{Si}{4} are often weaker 
(shallower) than, for example, \ion{C}{4} and \ion{O}{6}. 

The large line optical depths and possibly complex covering effects 
imply that we cannot derive elemental abundances from 
the BALs without a specific physical model. One constraint on such 
a model is that the partial line-of-sight 
coverage in PG~1254+047 applies specifically to the continuum source, 
because some of the strongest and most optically thick BALs 
(e.g. \ion{C}{4}) are far removed from any emission line.  
Therefore, either 1) there are physical structures 
in the BALR smaller than the projected size of the continuum 
source, which is conservatively $<$0.01~pc in the standard black 
hole-accretion disk paradigm (\cite{net92}), or 2) some of 
the observed continuum flux is scattered into our line of sight 
without passing through the BALR (\cite{coh95,goo95,hin95}).

Another important result for BALR models 
is that the high column densities inferred 
from the \ion{P}{5} line will produce significant absorption in soft X-rays 
(see, for example, the spectral simulations by \cite{shi95} and \cite{mur95b}). 
In fact, the minimum total column density derived here for the BALR, 
$\log N_{\rm H}$(\cmsq )~$\ga 22.0$ (Fig. 7), is nearly identical to the 
minimum absorbing columns inferred from the weak X-ray fluxes 
of other BALQSOs (\cite{gre96}). X-ray observations of PG~1254+047 
itself (\cite{wil94}) provide upper limits showing that this is also a weak X-ray 
source consistent with significant X-ray absorption. Its two-point spectral 
index between 2500~\AA\ and 2~keV is $\alpha_{ox}\leq -1.7$ 
($f_{\nu}\propto \nu^{\alpha_{ox}}$), 
compared to $\alpha_{ox}\leq -1.8$ for BALQSOs in general 
(\cite{gre96}). My calculations (Fig. 7) show that the column densities 
derived from the only two X-ray detections in BALQSOs, 
$\log N_{\rm H}$(\cmsq )~$\sim 23$ 
(\cite{mat95,gre96}), could produce the observed BALs in PG~1254+047 
if $+0.4\la \log U \la +0.7$ in a constant density medium. Therefore, the 
outflowing BALR is a strong candidate for the X-ray absorber. 
Furthermore, the profile similarities between \ion{P}{5} and strong lines 
such as \ion{C}{4}, \ion{N}{5} and \ion{O}{6} imply that the high column 
densities (and line saturation) occur across 
the entire range of absorption velocities, from $-$15,000 to 
$-$27,000~\kms . 

This result for high column densities at high velocities 
sets stringent requirements for the BALR acceleration. The equation 
of motion for a wind radiatively driven from a central point source with 
luminosity $L$ and mass $M$ is, 
\begin{equation}
{{v\ dv}\over{dR}} \ = \ {{f_L L}\over{4\pi R^2 c m_p N_{\rm H}}}\ - 
{{G M}\over{R^2}}
\end{equation}
where $v$ is the velocity, $R$ is the radial distance, and $f_L$ is the 
fraction of the total luminosity absorbed or scattered in the wind. 
Integrating Equation 2 from $R$ to infinity yields the terminal velocity, 
\begin{equation}
v_{\infty} \ \approx \ 32000\ R_{0.1}^{-{1\over 2}} 
\left({{{f_L L_{46}}\over{N_{22}}} - 
0.008 M_8}\right)^{{1}\over{2}} \ \ {\rm \kms}
\end{equation}
where $R_{0.1}$ is the inner wind radius in units of 0.1~pc, 
$L_{46}$ is the QSO luminosity in units of $10^{46}$~ergs~s$^{-1}$, 
$N_{22}$ is the total column density in $10^{22}$~\cmsq , and 
$M_8$ is the central black hole mass relative to 10$^8$~\Msun\ 
(see \cite{sco95} for similar expressions). $L_{46}\sim 1$ is the 
Eddington luminosity for $M_8 = 1$, and $R_{0.1}\sim 1$ is a 
nominal BEL region radius for $L_{46} = 1$ (cf. \cite{pet93}). 
Equations 2 and 3 hold strictly for 
open geometries, where the photons escaping one location in the 
BALR are not scattered or absorbed in another location. 
The radiative force from scattering in a single line 
is proportional to $1-e^{-\tau}$; therefore, optically thick lines 
cannot contribute more to the acceleration as the 
column densities (and optical depths) increase. However, at high 
column densities many more lines contribute. My line optical depth 
calculations for the case where $\tau_{\rm PV} = 0.2$ 
(\S6) indicate that the enemble BALs scatter $\sim$25\% of the 
continuum flux between 228 and 1600~\AA , near the peak of the 
QSO spectral energy distribution. 
Considering that there is also bound-free absorption due to 
helium and the metals, in addition to BALs at shorter wavelengths, 
it appears that $f_L$ could be as large as a few tenths. 
Therefore, according to Eqn. 3, the large total column densities 
implied by the \ion{P}{5} analysis (Fig. 7) {\it can} be radiatively 
driven to the observed high velocities. 
Radiative acceleration becomes even easier 
if the overall metallicity is above solar, because in that 
case the derived values of $N_{\rm H}$ would be lower. On the other 
hand, radiative driving could be problematic if the total column 
density is as high as $\log N_{\rm H}$(\cmsq )~$\sim 23$.

Another constraint emerges if we recast Eqn. 3 as a relationship 
between the velocity and the ionization parameter, namely, 
\begin{equation}
v_{\infty} \ \la\  50\left({{f_L U\overline{E_{Ryd}}\over{V_{fill}}}}\right)^{1\over 2} 
\ \ {\rm \kms}
\end{equation}
where $V_{fill}$ is the volume filling factor ($0\leq V_{fill}\leq 1$) and 
$\overline{E_{Ryd}}$ is the average luminosity-weighted photon energy 
in the Lyman continuum in units of Rydbergs. This relationship 
states simply that the momentum flux in the wind cannot exceed the 
momentum flux in photons (see also \cite{mur95b}). In conventional 
small-cloud models of the BALR (\cite{wey85,dek97}), Eqn. 4 shows that 
radiative acceleration to 
typical BALR velocities ($\ga$10,000~\kms ) can occur with 
ionization parameters of $U \la 1$ if $\log V_{fill} \la -4.6$. 
However, in the recent 
disk-wind model of Murray \etal (1995) and Murray \& Chiang (1996), 
$V_{fill}$ is unity and {\it extremely} high ionization parameters are 
required to drive the flow. In those models, an {\it additional} 
high-column density absorber is needed to shield the BALR from hard 
ionizing photons -- allowing moderate ionization species to exist at 
extremely high $U$ in the high-velocity BALR (see \cite{mur95b}). While it 
appears possible to radiatively drive a $V_{fill}\approx 1$ disk-wind with 
column densities as large as we infer from the 
\ion{P}{5} absorption (Fig. 7), that scenario requires another X-ray 
absorber (at low velocities) between the BALR and the continuum source. 
Murray \etal (1995) showed that the additional 
absorber must have $\log N_{\rm H}$(\cmsq )~$> 23$ and an ionization 
high enough to keep the gas optically thin at the Lyman limit. 

One final point. I noted in \S4 that the BALs in PG~1254+047 
are similar to existing measurements of other BALRs. We might,  
therefore, expect that \ion{P}{5} absorption and large column densities 
are typical of BALQSOs. More and better short-wavelength spectra 
are needed to search for \ion{P}{5} and test the saturation/high column 
density hypothesis in other sources. The theoretical line optical depths 
in Figure 6 provide a basis for these tests. Comparing the 
\ion{P}{5} lines to \ion{S}{5}, \ion{S}{6}, \ion{Ar}{5} or \ion{Ar}{6} 
transitions might be particularly useful because 1) these 
species have similar ionizations and 2) enrichment schemes that enhance 
the odd-numbered elements like phosphorus (\cite{shi96}) should produce 
large P/S and P/Ar abundance ratios (because S and Ar are even). 
For example, if the \ion{P}{5} detections {\it are} due to enhanced P 
abundances then \ion{Ar}{6} \lam 589 and \lam 755 should be absent, 
with optical depths $\la$0.01 times the combined \ion{P}{5} doublet 
for $\log U \la -1.0$ in the calculations of \S5.2. In contrast, the 
saturation/high column density hypothesis favored here predicts 
(Figure 6b) that these \ion{Ar}{6} lines should be $\ga$0.3 times the 
\ion{P}{5} pair. Unfortunately, these tests might be thwarted by line 
blending in most BALQSOs. 

\section{Summary}

I have used {\it HST} and ground-based spectra to study the 
\ion{P}{5}~\lam\lam 1118,1128 and other broad absorption lines 
(BALs) in the low-redshift QSO PG~1254+047. 
The \ion{P}{5} identification is secured by this line's 
redshift and profile similarity to other well-measured BALs, such as 
\ion{C}{4}~\lam\lam 1548,1550 and \ion{Si}{4}~\lam\lam 1393,1403, 
and by photoionization calculations showing that other 
lines near this wavelength (e.g. \ion{Fe}{3}~\lam 1123) should be much 
weaker than \ion{P}{5}. 

A standard analysis (\S5), which assumes that 
the line optical depths and ionic column densities can be derived from the 
residual intensities in the absorption troughs, leads to the following 
conclusions; 1) the ionization parameter in the photoionized BALR is 
$-1.8\la \log U \la -1.2$, 2) the total absorbing column is 
$18.4\la \log N_{\rm H}$(\cmsq )~$\la 19.0$, 3) the BALR is optically 
thin to UV and far-UV continuum radiation, 4) the BALs have optical depths 
ranging from a few tenths to a few, 6) the BALR has an extremely high 
metallicity, with [C/H]~$\ga +1.0$ and [Si/H]~$\ga +1.8$, and 
7) phosphorus is extremely overabundant, with [P/C]~$\ga +2.2$. 
These findings are typical of BALQSOs (\S1). The abundance results 
are particularly interesting because 
they are incompatible with a well-mixed interstellar medium 
enriched by a ``normal'' galactic stellar population 
(\S1 and \cite{ham97d}); more exotic enrichment schemes 
are needed (e.g. Shields 1996). 

However, the main result of this paper is that the BAL data are consistent 
with {\it solar} abundance ratios if at least some of the lines 
are more optically thick than they appear (\S6). In explicit photoionization 
calculations, the strength of the \ion{P}{5} BAL requires minimum line 
optical depths of $\ga$25 in \ion{C}{4}, $\ga$12 in \ion{N}{5} and 
$\ga$80 in \ion{O}{6} for solar abundance ratios. These minima apply 
for almost the entire range of BAL velocities, from $-$15,000 to 
$-$27,000~\kms . The minimum ionization parameter and total column 
density are $\log U\ga -0.6$ and $\log N_{\rm H}$(\cmsq )~$\ga 22.0$. 
In spite of the large column densities, the absence of low-ionization 
lines implies that the absorber is optically thin at the Lyman edge. 

I propose that the significant presence of \ion{P}{5} absorption 
is not an indicator of exotic abundances, but rather a byproduct of 
severe saturation in the other BALs. The large line optical depths 
are ``hidden'' in the observed (modest) BAL troughs because the absorber 
does not fully cover the background continuum source(s) along our 
line(s) of sight. Differences in the BAL strengths and profiles 
result from a complex mixture of optical depth, ionization 
and coverage fraction differences. This interpretation of the BALs in 
PG~1254+047 is supported by a variety of observational and theoretical 
studies of intrinsic absorption lines in other QSOs (\S1 and \S7). 

The large column densities inferred from the \ion{P}{5} BAL  
will produce significant bound-free absorption in soft X-rays. 
The outflowing BALR is therefore a strong candidate for the X-ray 
absorber in BALQSOs (\cite{mat95,gre96}). In particular, 
a nominal X-ray absorbing column of 
$\log N_{\rm H}$(\cmsq )~$\sim 23$ could produce the BAL spectrum 
in PG~1254+046 if $+0.4\la \log U \la +0.7$ in a single-zone medium. 
The predicted line optical depths in Figure 6 provide a basis for testing 
the high column density hypothesis with future UV spectral observations. 


\acknowledgments
I am grateful to T. A. Barlow and V. T. Junkkarinen for enlightening 
discussions, and to N. Murray, G. Shields, J. C. Shields and the referee 
R. Weymann for comments on this manuscript. I also thank G. J. Ferland 
for his generous support of the CLOUDY software. 
This work was funded by NASA grants NAG 5-1630 and NAG 5-3234.  

\clearpage
\begin{deluxetable}{lccc}
\tablewidth{0pt}
\tablecaption{Column Densities and Metal Abundances\tablenotemark{a}}
\tablecolumns{4}
\tablehead{
\colhead{Ion}& \colhead{log $N$}& 
\colhead{\ \ [M/H]$_p$}& \colhead{\ \ [M/H]$_{min}$}}
\startdata
\ion{H}{1}& \llap{$\la$}15.0& \nodata& \nodata\nl
\ion{C}{4}& 15.9& +1.4& +1.0\nl
\ion{Si}{4} \ \ & 15.0& +2.0& +1.8\nl
\ion{N}{5}& 16.2& +1.4& +1.0\nl
\ion{P}{5}& 15.0& +3.3& +3.1\nl
\enddata
\tablenotetext{a}{Metal abundances are 
[M/H]$_p$ for $f$(M$_i$) at their peaks and [M/H]$_{min}$ for the 
minimum $IC$. Column densities are in \cmsq . See \S5.1 and \S5.3.}\nl
\end{deluxetable}


\clearpage

\centerline{FIGURE CAPTIONS}
\bigskip

Fig. 1. --- {\it HST} spectrum of PG~1254+047 at the observed 
wavelengths. The Flux has units $10^{-15}$~\flam . 
The 1-$\sigma$ uncertainties 
are plotted across the bottom. The emission lines are labeled 
at $z_e = 1.010$ across the top. The wavelengths of possible 
BALs are labeled at 3 redshifts, $z_a = 0.855$, 0.871
and 0.893, corresponding to the 3 deepest minima in the \protect\ion{C}{4} 
trough. The label LL marks the position of the \protect\ion{H}{1} 
Lyman limit (912~\AA ). Not all of these labeled features are present. 
The smooth dotted curve is a fit to the continuum that 
includes a synthetic \protect\ion{O}{6} emission profile 
(see \S2 and \S3). 
\medskip

Fig. 2. --- 
Line profiles are plotted on a velocity scale relative to the emission 
redshift $z_e = 1.010$. The thin curve in each 
panel is the \protect\ion{C}{4} profile, while the thick curves are from 
top to bottom \protect\ion{Si}{4}, \protect\ion{N}{5}, 
\protect\ion{P}{5} and \protect\ion{O}{6}. 
The velocities of these doublets are defined by the 
equal-weight average of the individual rest wavelengths. 
The horizontal bar near the top-center of each panel shows the 
doublet separations. The \protect\ion{C}{4} separation 
(not shown) is negligible on this scale ($\sim$500~\kms ). 
Note that the plots often include unrelated emission 
or absorption lines (\S3). 
\medskip

Fig. 3. --- BAL optical depth profiles derived from the observed 
residual intensities using an algorithm that 
approximately removes the doublet structure (see \S5.1).
\medskip

Fig. 4. --- Theoretical line optical depths relative to 
\protect\ion{P}{5}~\lam 1118, $\tau /\tau(\lambda 1118)$, 
are plotted for solar abundances and different values of the ionization 
parameter, $U$, in optically thin clouds that are photoionized by a 
standard QSO continuum. 
The 3 panels show lines appearing at different wavelengths:  
$912 \leq \lambda < 1025$~\AA\ (left panel), 
$1025\leq \lambda < 1220$~\AA\ (middle) and 
$1220\leq\lambda < 3000$~\AA\ (right). The plots include all resonance 
lines at these wavelengths with the exceptions of the high-level 
\protect\ion{H}{1} lines and numerous 
low-ionization lines that would fill in the lower left corner of 
each panel.
The 2 thick horizontal curves in all 3 panels are the normalized 
optical depths in the \protect\ion{P}{5} doublet, \lam\lam 1118,1128. The 
absolute optical depths in the stronger \protect\ion{P}{5} transition, 
\lam 1118, are labeled across the top for $N_{\rm H} = 10^{20}$~\cmsq\ 
and a square absorption profile with FWHM~$= 10,000$~\kms . 
See \S3 and \S5.2. 
\medskip

Fig. 5. --- Logarithmic ionization corrections normalized to 
solar abundances, $IC$, are plotted for 
various ionization parameters, $U$, in calculations identical 
to Figure 4. The curve labeled 
\protect\ion{P}{5}/\protect\ion{C}{4} gives the correction 
factor for deriving the [P/C] abundance from Equation 1, etc.. The curve 
for \protect\ion{P}{5}/\protect\ion{Si}{4} is shifted up 
by +1 for convenience. 
\medskip

Fig. 6a. --- Theoretical line optical depths, $\tau$, for 
constant density, photoionized clouds that have solar abundances and 
square absorption profiles with FWHM~$= 10,000$~\kms . 
The total column density, $N_{\rm H}$ 
(labeled across the top), at each ionization parameter, $U$ (bottom), 
is adjusted so that the optical depth in \protect\ion{P}{5} 
(\lam 1118+\lam 1128) equals 0.2. The \protect\ion{P}{5} optical 
depths fall below this mark at $\log U = -1.0$ because no value 
of $N_{\rm H}$ can achieve $\tau_{\rm PV} = 0.2$ for $\log U\la -1.0$. 
The 3 panels show all resonance lines of significant strength between 912 
and 3000~\AA , as in Figure 4 (again excluding numerous 
low-ionization lines that would fill in the lower left corner of 
each panel). 
\medskip

Fig. 6b. --- Same as Figure 6a except for lines at shorter 
wavelengths: $584 \leq \lambda < 675$~\AA\ (left panel), 
$675\leq \lambda < 750$~\AA\ (middle) and 
$750\leq\lambda < 912$~\AA\ (right).
\medskip

Fig. 7. --- 
Column density limits for $N_{\rm H}$ (solid curves, left-hand scale) 
and $N_{\rm HI}$ (dash-dot curves, right-hand scale) for different 
ionization parameters, $U$, in 
constant density, photoionized clouds with solar abundances. 
The minimum $N_{\rm H}$ and $N_{\rm HI}$ values 
(lower curves for $\log U \ga -0.6$) derive from the requirement 
that $\tau_{PV} \geq 0.2$, while the maximum values 
(upper curves) follow from $\tau_{\rm MgII}$ and $\tau_{\rm AlIII} \leq 0.1$. 
The permitted values of $N_{\rm H}$ and $N_{\rm HI}$ lie between the two 
pairs of curves for $\log U \ga -0.6$, as indicated by the arrows (see 
\S6). 
\medskip

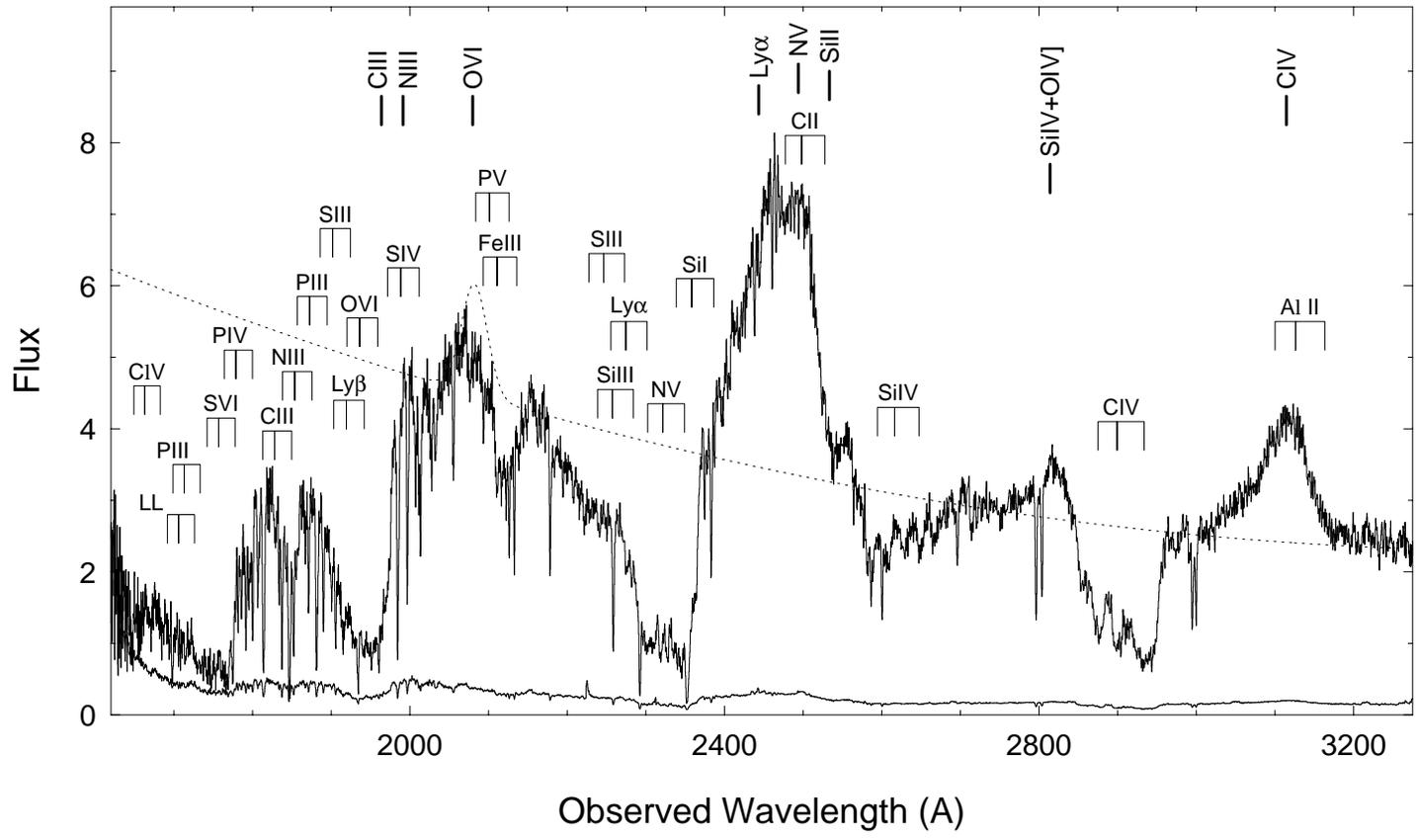
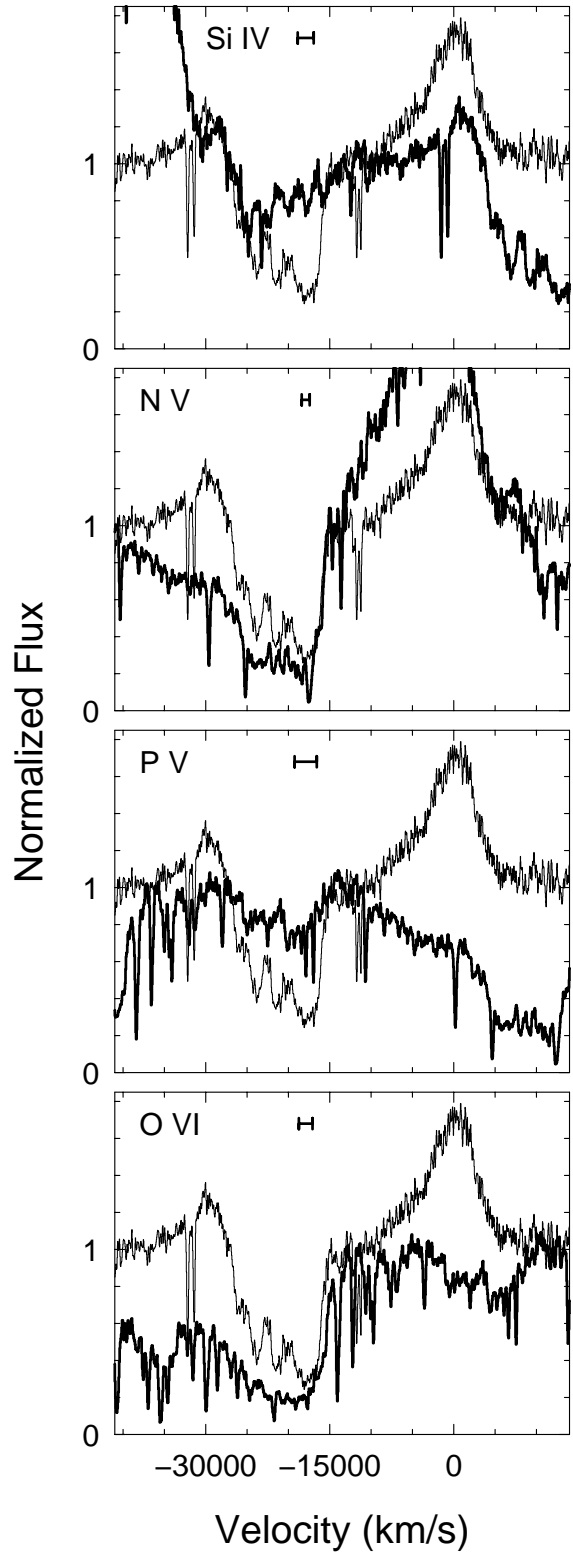
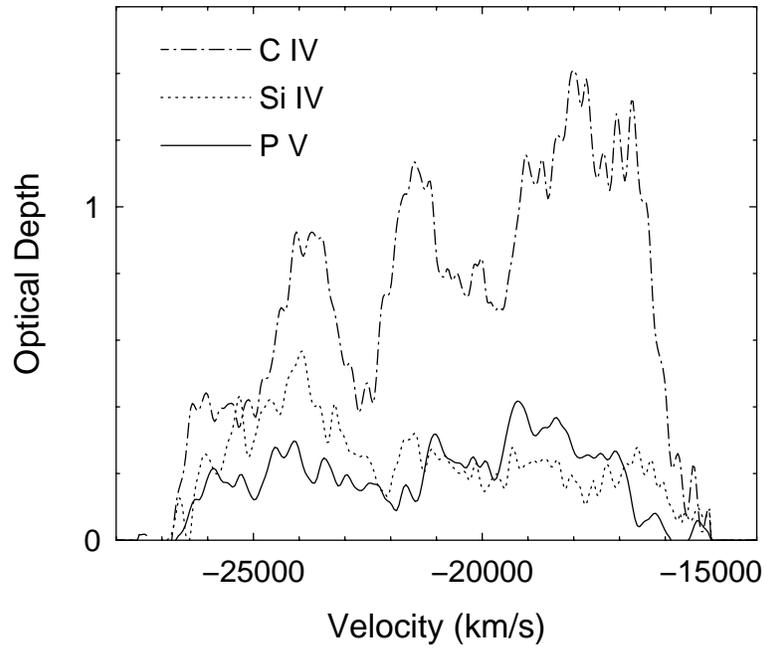
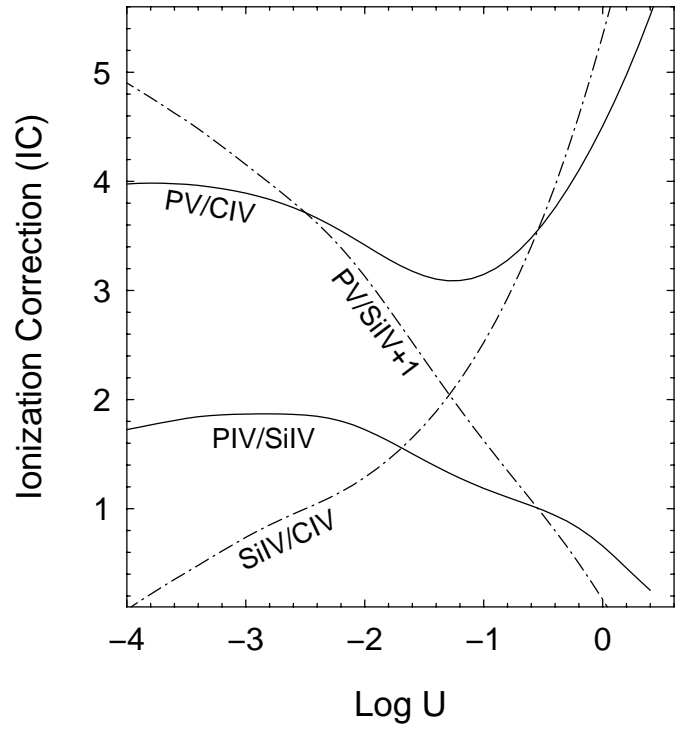
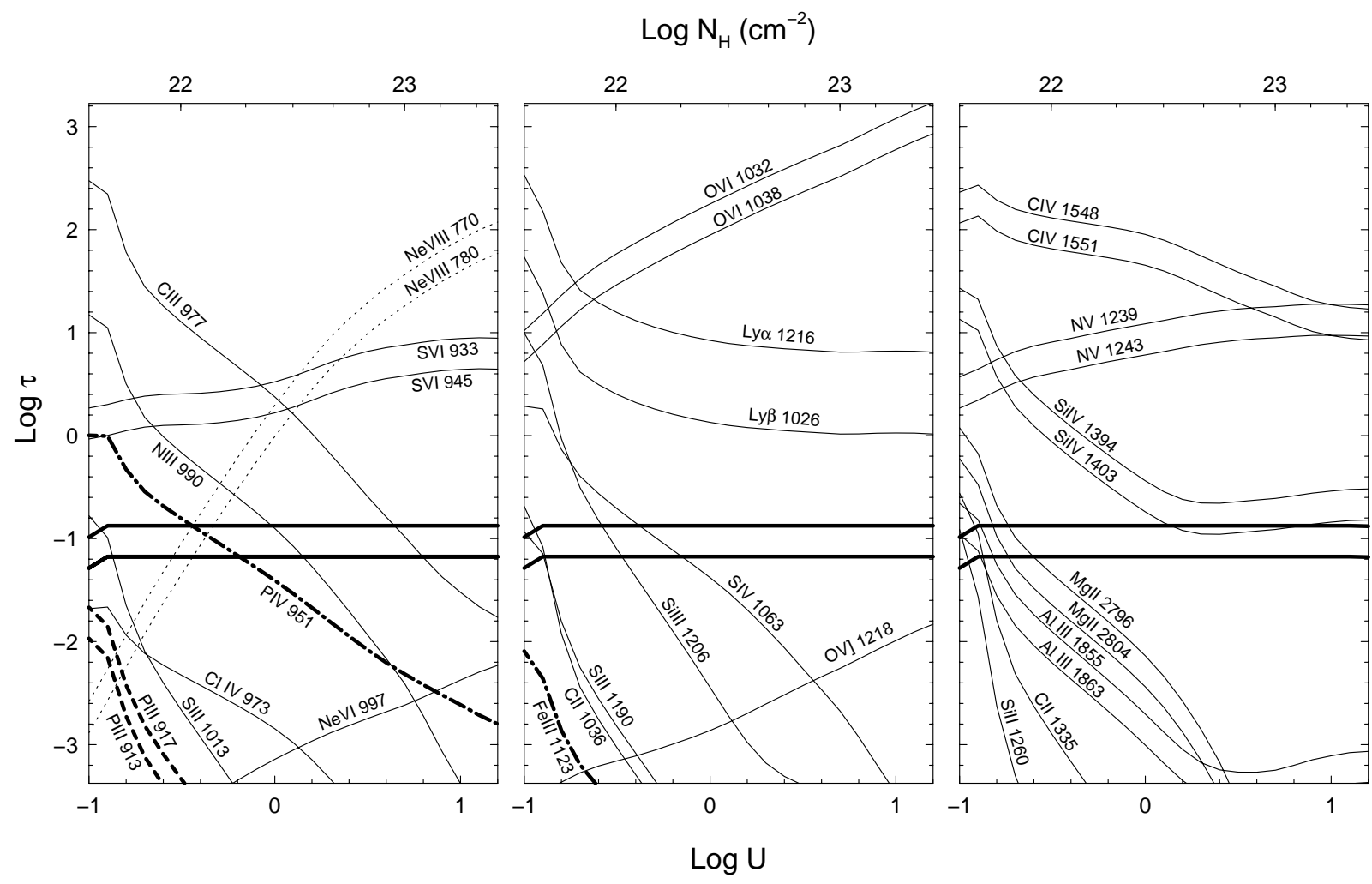
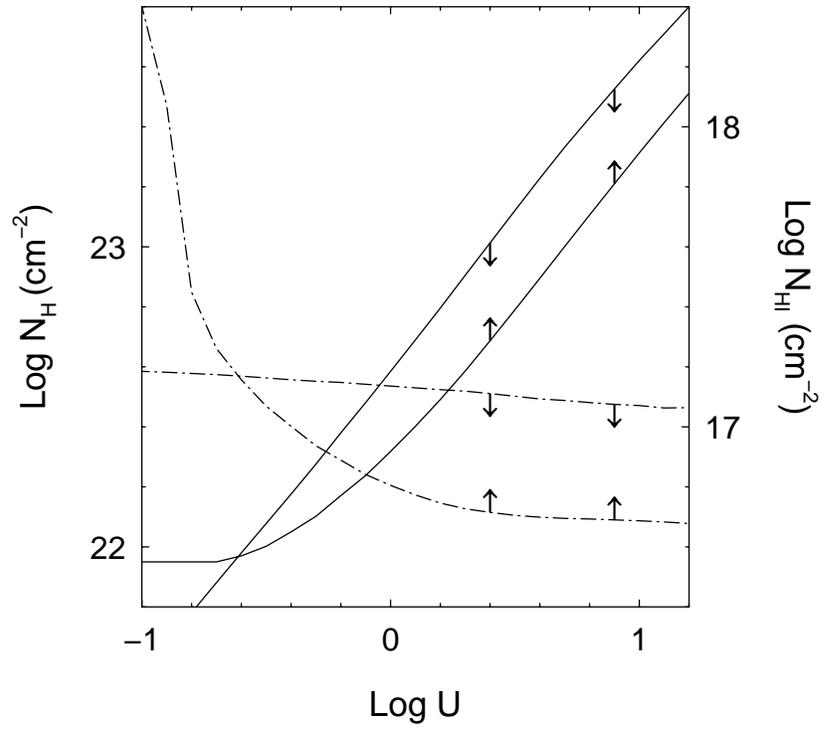
\end{document}